\documentclass[preprint,superscriptaddress,amssymb,amsmath,floatfix,citeautoscript]{revtex4-1}
\usepackage{soul}
\usepackage[pdftex]{graphicx}
\usepackage{amsmath}
\usepackage{natbib}
\usepackage{xcolor}

\begin{document}

\title{Experimental demonstration of a generalized Fourier's Law for non-diffusive thermal transport}

\author{Chengyun Hua\footnote{\text{
To whom correspondence should be addressed. E-mail: huac@ornl.gov; aminnich@caltech.edu}}}
\affiliation{Environmental and Transportation Science Division, Oak Ridge National Laboratory, Oak Ridge, TN 37831, USA}
\author{Lucas Lindsay}
\affiliation{Materials Science and Technology Division, Oak Ridge National Laboratory, Oak Ridge, TN 37831, USA}
\author{Xiangwen Chen}

\author{Austin J. Minnich$ ^{a}\ $}
\affiliation{%
Division of Engineering and Applied Science, California Institute of Technology, Pasadena, California 91125,USA
}

\date{\today}

\begin{abstract}

Phonon heat conduction over length scales comparable to their mean free paths is a topic of considerable interest for basic science and thermal management technologies. Although the failure of Fourier's law beyond the diffusive regime is well understood, debate exists over the proper physical description of thermal transport in the ballistic to diffusive crossover. Here, we derive a generalized Fourier's law that links the heat flux and temperature fields, valid from ballistic to diffusive regimes and for general geometries, using the Peierls-Boltzmann transport equation within the relaxation time approximation. This generalized Fourier's law predicts that thermal conductivity not only becomes nonlocal at length scales smaller than phonon mean free paths, but also requires the inclusion of an inhomogeneous nonlocal source term that has been previously neglected. We demonstrate the validity of this generalized Fourier's law through direct comparison with time-domain thermoreflectance (TDTR) measurements in the nondiffusive regime without adjustable parameters. Furthermore, we show that interpreting experimental data without this generalized Fourier's law leads to inaccurate measurement of thermal transport properties. 

\end{abstract}

\maketitle

\section{Introduction}

Fourier's law fails when a temperature gradient exists over a length scale comparable to or smaller than the mean free paths (MFPs) of heat carriers. In this regime, the heat flux and temperature fields may differ from the predictions of heat diffusion theory based on Fourier's law. These discrepancies have been observed at a localized hotspot created by a doped resistor thermometer in a suspended silicon membrane\cite{sverdrup_measurement_2001} and more recently in optical pump-probe experiments including soft x-ray diffraction from nanoline arrays,\cite{highland_ballistic-phonon_2007, siemens_quasi-ballistic_2010} transient grating,\cite{johnson_direct_2013} and thermoreflectance methods.\cite{cahill_nanoscale_2014, koh_frequency_2007, minnich_thermal_2011, regner_broadband_2012, english_mean_2013, hu_spectral_2015, Hua_2017, ravichandran_spectrally_2018} In particular, due to the absence of scattering the transport properties become nonlocal, in contrast to Fourier's law in which the heat flux at a certain location is determined by the temperature gradient only at that location. 

Lattice thermal transport in crystals is generally described by the Peierls-Boltzmann equation (PBE), first derived by Peierls,\cite{Peierls1929} from which the thermal conductivity is given in terms of the microscopic properties of phonons.\cite{Peierls1929,Peierls1955} However, solving the PBE for a general space-time dependent problem remains a challenging task due to the high dimensionality of the integro-differential equation.  Thus, most prior works have determined solutions of the PBE in certain limiting cases. 

Guyer and Krumhansl\cite{guyer_solution_1966} first performed a linear response analysis of the PBE, deriving a space-time-dependent thermal conductivity by assuming the Normal scattering rates were much larger than Umklapp scattering rates, and they applied their solution to develop a phenomenological coupling between phonons and elastic dilatational fields caused by lattice anharmonicity. Hardy and coworkers reported a rigorous quantum-mechanical formulation of the theory of lattice thermal conductivity using a perturbation method that included both anharmonic forces and lattice imperfections.\cite{hardy_energy-flux_1963,hardy_perturbation_1965, hardy_lowestorder_1965} This quantum treatment of lattice dynamics was then verified by both theoretically and experimentally demonstrating the presence of Poiseuille flow and the second sound in a phonon gas at low temperatures when Umklapp processes may be neglected.\cite{sussmann_thermal_1963, guyer_thermal_1966, hardy_phonon_1970, jackson_thermal_1971,beck_phonon_1974} The variational principle was also used to solve the PBE with Umklapp scattering incorporated.\cite{hamilton_variational_1969, srivastava_derivation_1976} Levinson developed a nonlocal diffusion theory of thermal conductivity from a solution of the PBE with three-phonon scattering in the low frequency limit.\cite{levinson_nonlocal_1980} 

Advances in computing power have enabled the numerical solution of the PBE with inputs from density functional theory, fully \emph{ab initio}. For instance, bulk lattice thermal conductivities are now routinely computed from first principles using an iterative solution of the PBE\cite{ward_ab_2009,broido_lattice_2005,li_shengbte:_2014,carrete_almabte_2017,omini_iterative_1995} or from variational approaches.\cite{fugallo_ab_2013} Chaput\cite{chaput_direct_2013} presented a direct solution to the transient linearized PBE with an imposed constant temperature gradient. Cepellotti and Marzari\cite{cepellotti_thermal_2016} introduced the concept of a "relaxon", an eigenstate of the symmetrized scattering operator of the PBE first used by Guyer \emph{et. al.}\cite{guyer_solution_1966} and Hardy\cite{hardy_phonon_1970}. They applied this treatment to solve steady-state problems in two-dimensional systems with a constant temperature gradient imposed in one direction\cite{cepellotti_boltzmann_2017}.

Solving the PBE with the full collision operator, even in its linearized form, is difficult for complicated geometries. Therefore, various theoretical frameworks based on a simplified PBE have been developed to describe nonlocal thermal transport for general problems. Non-diffusive responses observed in experiments\cite{johnson_direct_2013, cahill_nanoscale_2014, koh_frequency_2007, regner_broadband_2012, Wilson2013,Dames2015} have been explained using the phonon-hydrodynamic heat equation\cite{torres_emergence_2018}, a truncated Levy formalism\cite{Vermeersch2015b}, a two-channel model in which low and high frequency phonons are described by the PBE and heat equation\cite{Maznev2011}, and a Mckelvey-Shockley flux method\cite{Maassen2015}. Methods based on solving the PBE under the relaxation time approximation (RTA), where each phonon mode relaxes towards thermal equilibrium at a characteristic relaxation rate, have been developed to investigate nonlocal transport in an infinite domain\cite{Mahan_1988,hua_analytical_2014,allen_temperature_2018,Collins2013APL}, a finite one-dimensional slab\cite{Hua_semi_analytical_2015, Koh_2014}, and experimental configurations such as transient grating\cite{Ramu2014,Collins2013APL} and thermoreflectance experiments\cite{Regner2014,zeng_disparate_2014, Vermeersch_2015a}. An efficient Monte Carlo scheme has been used to solve the PBE under the RTA in complicated geometries involving multiple boundaries\cite{Peraud:2011,Peraud:2012,hua_importance_2014}.

In the diffusion regime, Fourier's law is a relation between heat flux and temperature fields, independent of the specific problem. In the nondiffusive regime, obtaining such a relation is more complicated because the transport is inherently nonlocal. The works described above solve the PBE for problems with specific boundary conditions or inputs. Thus, despite these efforts, a description generalizing Fourier's law for arbitrary geometries and linking the heat flux and temperature fields in all transport regimes is not available. 

Here, we derive a generalized Fourier's law to describe non-diffusive thermal transport for general geometries using the linearized PBE within the RTA. The generalized Fourier's law requires the inclusion of an inhomogeneous nonlocal term arising from the source or the boundary conditions of the particular problem. By including the inhomogeneous contribution to the heat flux, the space- and time-dependent thermal conductivity is independent of the specific geometry or inputs. This generalized Fourier's law is validated by favorable comparisons with a series of TDTR measurements in the non-diffusive regime. We also show that neglecting the inhomogeneous contribution to the heat flux leads to inaccurate measurement of thermal transport properties in the non-diffusive regime. Our work provides a unified description of heat transport for a wide range of problems from ballistic to diffusive regimes. 

\section{Theory}

\subsection{Governing Equation}
We begin by briefly reviewing the derivation of the transport solution to the PBE. The mode-dependent PBE under the relaxation time approximation for transport is given by
\begin{equation}\label{eq:PBTE}
\frac{\partial g_{\mu}(\mathbf{x},t)}{\partial t}+\mathbf{v}_{\mu}\cdot\nabla g_{\mu}(\mathbf{x},t)= - \frac{g_{\mu}(\mathbf{x},t)-g_{0}(T(\mathbf{x},t))}{\tau_{\mu}}+\dot{Q}_{\mu}(\mathbf{x},t),
\end{equation}
where $g_{\mu}(\mathbf{x},t) = \hbar\omega_{\mu}(f_{\mu}(\mathbf{x},t)-f_0(T_0))$ is the deviational energy distribution function at position $\mathbf{x}$ and time $t$ for phonon states $\mu$ ($\mu \equiv (\mathbf{q},s)$, where $\mathbf{q}$ is the wavevector and $s$ is the phonon branch index). $f_0$ is the equilibrium Bose-Einstein distribution, and $g_{0}(T(\mathbf{x},t)) = \hbar\omega_{\mu}(f_0(T(\mathbf{x},t))-f_0(T_0)) \approx C_{\mu}\Delta T(\mathbf{x},t)$, where $T(\mathbf{x},t)$ is the local temperature, $T_0$ is the global equilibrium temperature, $\Delta T(\mathbf{x},t) = T(\mathbf{x},t)-T_0$ is the local temperature deviation from the equilibrium value, and $C_{\mu}=\hbar\omega_{\mu}\frac{\partial f_0}{\partial T}|_{T_0}$ is the mode-dependent specific heat. Here, we assume that $\Delta T(\mathbf{x},t)$ is small such that $g_{0}(T(\mathbf{x},t))$ is approximated to be the first term of its Taylor expansion around $T_0$. Finally, $\dot{Q}_{\mu}(\mathbf{x},t)$ is the heat input rate per mode, $\mathbf{v}_{\mu} = (v_{\mu x}, v_{\mu y},v_{\mu z})$ is the phonon group velocity vector, and $\tau_{\mu}$ is the phonon relaxation time. 

To close the problem, energy conservation is used to relate $g_{\mu}(\mathbf{x},t)$ to $\Delta T(\mathbf{x},t)$ as
\begin{equation}\label{eq:EnergyConservation_General}
\frac{\partial E(\mathbf{x},t)}{\partial t}+\nabla \cdot \mathbf{q}(\mathbf{x},t) = \dot{Q}(\mathbf{x},t),
\end{equation}
where $E(\mathbf{x},t) =V^{-1}\sum_{\mu}g_{\mu}(\mathbf{x},t)$ is the total volumetric energy, $\mathbf{q}(\mathbf{x},t)= V^{-1} \sum_{\mu} g_{\mu}(\mathbf{x},t)\mathbf{v_{\mu}}$ is the directional heat flux, and $\dot{Q}(\mathbf{x},t) = V^{-1}\sum_{\mu}\dot{Q}_{\mu}(\mathbf{x},t)$ is the volumetric mode-specific heat input rate. Here, the sum over $\mu$ denotes a sum over all phonon modes in the Brillouin zone, and $V$ is the volume of the crystal. The solution of Eq.~(\ref{eq:PBTE}) yields a distribution function, $g_{\mu}(\mathbf{x},t)$, from which temperature and heat flux fields can be obtained using Eq.~(\ref{eq:EnergyConservation_General}). Like the classical diffusion case, the exact expression of the temperature field varies from problem to problem. However, in a diffusion problem, the constitutive law that links the temperature and heat flux fields is governed by one expression, Fourier's law. Here, we seek to identify a similar relation that directly links temperature and heat flux fields for non-diffusive transport, regardless of the specific problem. 

To obtain this relation, we begin by rearranging Eq.~(\ref{eq:PBTE}) and performing a Fourier transform in time $t$ on Eq.~(\ref{eq:PBTE}), which gives
\begin{equation}\label{eq:linearizedBTE}
\Lambda_{\mu x}\frac{\partial \tilde{g}_{\mu}}{\partial x} + \Lambda_{\mu y}\frac{\partial \tilde{g}_{\mu}}{\partial y}+\Lambda_{\mu z}\frac{\partial \tilde{g}_{\mu}}{\partial z}+(1+i\eta\tau_{\mu}) \tilde{g}_{\mu} =C_{\mu} \Delta \tilde{T}+\tilde{Q}_{\mu}\tau_{\mu}, 
\end{equation}
where $\eta$ is the Fourier temporal frequency, and $\Lambda_{\mu x}$, $\Lambda_{\mu y}$ and $\Lambda_{\mu z}$ are the directional mean free paths along $x$, $y$, and $z$ directions, respectively. Equation~(\ref{eq:linearizedBTE}) can be solved by defining a new set of independent variables $\xi$, $\rho$, and $\zeta$ such that
\begin{subequations}\label{eq:newvariables}
\begin{eqnarray}
\xi &=& x, \label{eq:xi}\\
\rho &=& \frac{\Lambda_{\mu y}}{\Lambda_{\mu} }x-\frac{\Lambda_{\mu x}}{\Lambda_{\mu} }y, \label{eq:eta}\\
\zeta &=&\frac{\Lambda_{\mu z}}{\Lambda_{\mu} }x-\frac{\Lambda_{\mu x}}{\Lambda_{\mu} }z,
\end{eqnarray}
\end{subequations}
where $\Lambda_{\mu} =\sqrt{\Lambda^2_{\mu x}+\Lambda^2_{\mu y}+\Lambda^2_{\mu z}}$. The Jacobian of this transformation is $\Lambda^2_{\mu x}/\Lambda^2_{\mu}$, a nonzero value. After changing the coordinates from $(x,y,z)$ to the new coordinate system $(\xi,\rho,\zeta)$, $(v_{\mu \xi} = v_{\mu x}, 0, 0)$ is the set of elements for the velocity vector $\mathbf{v}_{\mu}$ in the new coordinates, and Eq.~(\ref{eq:linearizedBTE}) becomes a first order partial differential equation with only one partial derivative
\begin{equation}\label{eq:BTENewCoordinate}
\Lambda_{\mu \xi} \frac{\partial \tilde{g}_{\mu}}{\partial \xi} +\alpha_{\mu}\tilde{g}_{\mu} = C_{\mu} \Delta \tilde{T}+\tilde{Q}_{\mu}\tau_{\mu},
\end{equation}
where $\alpha_{\mu} = 1 + i\eta\tau_{\mu}$.
Assuming that $\xi \in [L_1, L_2]$, Eq.~(\ref{eq:BTENewCoordinate}) has the following solution:
\begin{subequations}
\begin{eqnarray}
\tilde{g}_{\mu}(\xi,\rho,\zeta,\eta) &=& P^{+}_{\mu}e^{-\alpha_{\mu}\frac{\xi-L_1}{\Lambda_{\mu \xi} }}+\int^{\xi}_{L_1}\frac{C_{\mu} \Delta \tilde{T}+\tilde{Q}_{\mu}\tau_{\mu}}{\Lambda_{\mu \xi} }e^{-\alpha_{\mu}\frac{\xi-\xi '}{\Lambda_{\mu \xi} }}d\xi ' \text{ for } v_{\mu \xi} > 0, \label{eq:distfunction_plus}\\
\tilde{g}_{\mu}(\xi,\rho,\zeta,\eta)&=& P^{-}_{\mu}e^{\alpha_{\mu}\frac{L_2-\xi}{\Lambda_{\mu \xi} }}-\int^{L_2}_{\xi}\frac{C_{\mu} \Delta \tilde{T}+\tilde{Q}_{\mu}\tau_{\mu}}{\Lambda_{\mu \xi} }e^{-\alpha_{\mu}\frac{\xi-\xi'}{\Lambda_{\mu \xi} }}d\xi '  \text{ for } v_{\mu \xi} <0. \label{eq:distfunction_minus}
\end{eqnarray}
\end{subequations}
$P^{+}_{\mu}$ and $P^{-}_{\mu}$ are functions of $\rho$, $\zeta$, $\eta$ and are determined by the boundary conditions at $\xi = L_1$ and $\xi = L_2$, respectively. Using the symmetry of $v_{\mu \xi}$ about the center of the Brillouin zone, i.e., $v_{\mu\xi} = -v_{-\mu\xi}$, Eqs.~(\ref{eq:distfunction_plus}) \& (\ref{eq:distfunction_minus}) can be combined into the following form: 
\begin{equation}\label{eq:distfunction}
\tilde{g}_{\mu}(\xi,\rho,\zeta,\eta)= P_{\mu}e^{-\alpha_{\mu}\frac{\xi}{\Lambda_{\mu \xi}}}+\int_{\Gamma}\frac{C_{\mu} \Delta \tilde{T}+\tilde{Q}_{\mu}\tau_{\mu}}{|\Lambda_{\mu \xi}| }e^{-\alpha_{\mu}\left|\frac{\xi'-\xi }{\Lambda_{\mu \xi} }\right|}d\xi ',
\end{equation}
where 
\begin{equation}\label{eq:BoundaryConditions}
P_{\mu} = \left\{ \begin{array}{rl}
P^{+}_{\mu}e^{\alpha_{\mu}\frac{L_1}{\Lambda_{\mu \xi} }} &\mbox{ if $v_{\mu\xi}>0$} \\
P^{-}_{\mu}e^{\alpha_{\mu}\frac{L_2}{\Lambda_{\mu \xi} }} &\mbox{ if $v_{\mu\xi}<0$}
\end{array} \right.
\end{equation}
and 
\begin{equation}\label{eq:Limits}
\Gamma \in \left\{ \begin{array}{rl}
 [L_1, \xi ) &\mbox{ if $v_{\mu\xi}>0$} \\
 \left(\xi , L_2 \right] &\mbox{ if $v_{\mu\xi}<0$}
\end{array} \right. .
\end{equation}
The energy conservation equation becomes
\begin{equation}\label{eq:EnergyConservation}
\sum_{\mu}v_{\mu \xi}\frac{\partial \tilde{g}_{\mu}}{\partial \xi}+i\eta\sum_{\mu}\tilde{g}_{\mu} = \sum_{\mu}\tilde{Q}_{\mu},
\end{equation}
where $v_{\mu \xi} \tilde{g}_{\mu}$ gives the mode-specific heat flux along the $\xi$ direction expressed as:
\begin{equation}\label{eq:HeatFlux}
\tilde{q}_{\mu\xi}= P_{\mu}v_{\mu \xi}e^{-\alpha_{\mu}\frac{\xi}{\Lambda_{\mu \xi} }} +\int_{\Gamma}\tilde{Q}_{\mu}(\xi',\rho,\zeta,\eta)e^{-\alpha_{\mu}\left|\frac{\xi-\xi '}{\Lambda_{\mu \xi} }\right|}d\xi ' + \int_{\Gamma}\frac{C_{\mu}v_{\mu \xi}}{|\Lambda_{\mu \xi}|} \Delta \tilde{T}(\xi ',\rho,\zeta,\eta)e^{-\alpha_{\mu}\left|\frac{\xi-\xi '}{\Lambda_{\mu \xi} }\right|}d\xi '.
\end{equation}

Applying integration by parts to the third term in Eq.~(\ref{eq:HeatFlux}), we can write the heat flux per mode as:
\begin{equation}\label{eq:HeatFlux_v2}
\tilde{q}_{\mu \xi} = - \int_{\Gamma}\kappa_{\mu\xi}(\xi-\xi')\frac{\partial \tilde{T}}{\partial \xi'}d\xi '+B_{\mu}(\xi,\rho,\zeta,\eta),
\end{equation}
where 
\begin{eqnarray}
B_{\mu}(\xi,\rho,\zeta,\eta) &=& P_{\mu}v_{\mu\xi}e^{-\alpha_{\mu}\frac{\xi}{\Lambda_{\mu \xi}}}+\frac{C_{\mu}|v_{\mu\xi}|}{\alpha_{\mu}}e^{-\alpha_{\mu}\frac{\xi}{\Lambda_{\mu \xi} }}\left[\Delta \tilde{T}e^{\alpha_{\mu}\frac{\xi'}{\Lambda_{\mu \xi} }}\right]_{\Gamma} \nonumber \\
&+&sgn(v_{\mu \xi}) \int_{\Gamma}\tilde{Q}_{\mu}(\xi',\rho,\zeta,\eta)e^{-\alpha_{\mu}\left|\frac{\xi-\xi '}{\Lambda_{\mu \xi}}\right|}d\xi ',
\end{eqnarray}
is solely determined by the boundary condition and the volumetric heat input rate.
$\kappa_{\mu\xi}(\xi)$ is the modal thermal conductivity along the $\xi$ direction given by 
\begin{equation}
\kappa_{\mu\xi}(\xi) =C_{\mu}v_{\mu\xi}\Lambda_{\mu\xi}\frac{e^{-\alpha_{\mu}\left|\frac{\xi}{\Lambda_{\mu \xi}}\right|}}{\alpha_{\mu}|\Lambda_{\mu \xi}|}.
\end{equation}

Equation (\ref{eq:HeatFlux_v2}) is the primary result of this work. This equation links temperature gradient to the mode-specific heat flux for a general geometry. Since this constitutive equation of heat conduction is valid from ballistic to diffusive regimes, we denote it as a generalized Fourier's law. It describes that for a specific phonon mode $\mu$,  heat only flows in the $\xi$ direction in the new coordinate system ($\xi$, $\rho$, $\zeta$) since the velocities in $\rho$ and $\zeta$ directions are zero. To obtain the total heat flux in the original coordinate system, \emph{e.g.} $q_x$, $q_y$, and $q_z$ in Cartesian coordinates, all the functions involved in Eq.~(\ref{eq:HeatFlux_v2}) must be mapped from the coordinate system ($\xi$, $\rho$, $\zeta$) to ($x$, $y$, $z$). Analytical mappings exist for a few special cases that we will discuss shortly. 

There are two parts in Eq.~(\ref{eq:HeatFlux_v2}). The first part represents a convolution between the temperature gradient along the $\xi$ direction and a space-and time-dependent thermal conductivity, $\kappa_{\mu\xi}(\xi)$. As reported previously, this convolution indicates the nonlocality of the thermal conductivity.\cite{Mahan_1988, Koh_2014, allen_temperature_2018} However, a second term exists that is determined by the inhomogeneous term originating from the boundary conditions and source terms. Similar to the first term, the contribution from the heat input to the heat flux, given by $\int_{\Gamma}\tilde{Q}_{\mu}(\xi',\rho,\zeta,\eta)e^{-\alpha_{\mu}\left|\frac{\xi-\xi'}{\Lambda_{\mu \xi}}\right|}d\xi $, is nonlocal, meaning the contribution at a given point is determined by convolving the heat source function with an exponential decay function with a decay length of $\Lambda_{\mu x}$. 

While the nonlocality of thermal conductivity was identified by earlier works on phonon transport\cite{guyer_solution_1966,guyer_thermal_1966,levinson_nonlocal_1980,koh_frequency_2007,Vermeersch2015b,Mahan_1988,Allen_2018,allen_temperature_2018}, the contribution from the inhomogeneous term has been neglected. Recently, Allen and Perebeinos\cite{allen_temperature_2018} considered the effects of external heating and derived a thermal susceptibility based on the PBE that links external heat input to temperature response and a thermal conductivity that links temperature response to heat flux. However, their derived thermal susceptibilities and thermal conductivities are subject to the specific choice of the external heat input. In this work, we demonstrate that there exists a general relation between heat flux and temperature distribution without specifying the geometry of the problem. The space- and time-dependent thermal conductivity in the first term of Eq.~(\ref{eq:HeatFlux_v2}) is independent of boundary conditions and heat input. The dependence of heat flux on the specific problem is accounted for by the inhomogeneous term.

\subsection{Diffusive limit}

Here we examine some specific limits of Eq.~(\ref{eq:HeatFlux_v2}). First, in the diffusive regime, the spatial and temporal dependence of thermal conductivity disappears and asymptotically approaches a constant. To demonstrate this limit, we first identify two key nondimensional parameters in Eq.~(\ref{eq:HeatFlux_v2}): Knudsen number, Kn$_{\mu}$ $\equiv \Lambda_{\mu \xi}\xi^{-1}$, which compares phonon MFP with a characteristic length, in this case $\xi^{-1}$, and a transient number, $\Xi_{\mu} \equiv \eta\tau_{\mu} $, which compares the phonon relaxation times with a characteristic time, in this case $\eta^{-1}$. In the diffusive limit, both $\Xi$ and Kn are much less than unity. Then, Eq.~(\ref{eq:distfunction}) is simplified to $C_{\mu}\Delta \tilde{T}$, and Eq.~(\ref{eq:HeatFlux_v2}) becomes
\begin{equation}
\tilde{q}_{\mu\xi} =  - \int_{\Gamma}C_{\mu}v_{\mu\xi}\Lambda_{\mu\xi}\frac{\partial \tilde{T}}{\partial \xi'}\delta(\xi-\xi')d\xi ' = - \kappa_{\mu \xi}\frac{\partial \tilde{T}}{\partial \xi},
\end{equation}
since in this limit we can perform the following simplifications
\begin{equation}
\lim_{\Xi \rightarrow 0} \alpha_{\mu} \approx 1,
\end{equation}
\begin{equation}
\lim_{\Xi, \text{ Kn}\rightarrow 0} e^{-\alpha_{\mu}\frac{\xi}{|\Lambda_{\mu \xi}| }} \approx 0,
\end{equation}
\begin{equation}
\lim_{\Xi, \text{ Kn}\rightarrow 0} \frac{e^{-\alpha_{\mu}\left|\frac{\xi-\xi'}{\Lambda_{\mu \xi}}\right|}}{|\Lambda_{\mu x}|} \approx \delta(\xi-\xi').
\end{equation}
The equation of energy conservation becomes
\begin{equation}\label{eq:EC_diffusive}
-\sum_{\mu} \kappa_{\mu \xi}\frac{\partial^2 \tilde{T}}{\partial \xi^2}+i\eta\sum_{\mu}C_{\mu}\Delta \tilde{T}= \sum_{\mu} \tilde{Q}_{\mu}. 
\end{equation}
Since $\frac{\partial}{\partial \xi} = \frac{\partial}{\partial x}+\frac{\partial}{\partial y}\frac{\Lambda_{\mu y}}{\Lambda_{\mu x}}+\frac{\partial}{\partial z}\frac{\Lambda_{\mu z}}{\Lambda_{\mu x}}$, Eq.~(\ref{eq:EC_diffusive}) can be mapped back to Cartesian coordinates as
\begin{equation}\label{eq:heatdiffusion}
\kappa_{x}\frac{\partial^2 \tilde{T}}{\partial x^2}+\kappa_{y}\frac{\partial^2 \tilde{T}}{\partial y^2}+\kappa_{z}\frac{\partial^2 \tilde{T}}{\partial z^2}+i\eta\Delta \tilde{T} \sum_{\mu}C_{\mu}= \sum_{\mu}\tilde{Q}_{\mu},
\end{equation}
where $\kappa_{i} = \sum_{\mu}C_{\mu}v_{\mu i}\Lambda_{\mu i}$ is the thermal conductivity along axis $i = x, y$, or $z$. Here, we assume that the off-diagonal elements of the thermal conductivity tensor are zero, i.e., $\kappa_{ij} = \sum_{\mu}C_{\mu}v_{\mu i}\Lambda_{\mu j} = 0$ when $i \neq  j$. Equation ~(\ref{eq:heatdiffusion}) is the classical heat diffusion equation.

\subsection{Generalized Fourier's law in a transient grating experiment}

We now check another special case of Eq.~(\ref{eq:HeatFlux_v2}) by applying it to the geometry of a one-dimensional transient grating experiment.\cite{johnson_direct_2013, Hua_transport_2014} Since it is a 1D problem, $\xi$ in Eq.~(\ref{eq:HeatFlux_v2}) is equivalent to $x$. In this experiment, the heat input has a spatial profile of $e^{i\beta x}$ in an infinite domain, where $\beta \equiv 2\pi/L$ and $L$ is the grating period. The boundary term vanishes, i.e., $\xi \in (-\infty, \infty)$, and both the distribution function and temperature field exhibit the same spatial dependence.  Then, the total heat flux is expressed as
\begin{equation}\label{eq:TGHeatFlux}
\tilde{q}_{x}(x,\eta) = i\beta\tilde{T}(\eta)e^{i\beta x}\sum_{\mu x>0}\frac{\kappa_{\mu x}}{\alpha_{\mu}^2+\Lambda_{\mu x}^2\beta^2}+\sum_{\mu x >0}\frac{Q_{\mu}}{\delta}\frac{e^{i\beta x}\alpha_{\mu}\Lambda_{\mu x}}{\alpha_{\mu}^2+\Lambda_{\mu x}^2\beta^2},
\end{equation}
where the total volumetric energy deposited on a sample is given by $\sum_{\mu}Q_{\mu}$, and the duration of the energy deposition is $\delta$. A derivation of Eq.~(\ref{eq:TGHeatFlux}) is given in Appendix \ref{app:TGDerivation}. 

The time scale of a typical TG experiment is on the order of a few hundred nanoseconds while relaxation times of phonons are typically less than a nanosecond for many semiconductors at room temperature. Therefore, we assume that $\Xi\ll 1$, and Eq.~(\ref{eq:TGHeatFlux}) is simplified to
\begin{equation}\label{eq:TGHeatFlux_simplified}
\tilde{q}_{x}(x,\eta) = i\beta\tilde{T}(\eta)e^{i\beta x}\sum_{\mu_x>0}\frac{\kappa_{\mu x}}{1+\Lambda_{\mu x}^2\beta^2}+\sum_{\mu_x>0}\frac{Q_{\mu}}{\delta}\frac{e^{i\beta x}\Lambda_{\mu x}}{1+\Lambda_{\mu x}^2\beta^2}.
\end{equation}
which is consistent with what has been derived in our earlier work.\cite{Hua_transport_2014,Maznev2011} The first part of Eq.~(\ref{eq:TGHeatFlux_simplified}) represents the conventional understanding of nonlocal thermal transport, a Fourier type relation with a reduced thermal conductivity given by 
\begin{equation}
\kappa_x = \sum_{\mu_x >0}\frac{\kappa_{\mu x}}{1+\Lambda_{\mu x}^2\beta^2},
\end{equation}
while the second part of the equation represents the contribution from the heat source to the total heat flux, which increases as the Knudsen number $\Lambda_{\mu x}/L$ increases. In a TG experiment, the presence of a single spatial frequency simplifies the convolutions in Eq.~(\ref{eq:HeatFlux_v2}) into products, and the only time dependence comes from the temperature. Therefore, the decay rate of the measured transient temperature profile is directly proportional to the reduced thermal conductivity. In general, the spatial dependency of the temperature field is less complicated in a TG experiment than in other experiments, making the separation of the intrinsic thermal conductivity contribution from the inhomogeneous contribution easier. 

\subsection{Generalized Fourier's law with infinite transverse geometries}
The third special case considered here is when the $y$ and $z$ directions extend to infinity. The analytical mapping of Eq.~(\ref{eq:HeatFlux_v2}) to Cartesian coordinates can be completed via Fourier transform in $y$ and $z$. After Fourier transform, Eq.~(\ref{eq:HeatFlux_v2}) becomes
\begin{equation}\label{eq:heatflux_3D}
\tilde{q}_{x}(x,f_y,f_z,\eta) = -\int^{L_2}_{L_1}\kappa_x(x-x',f_y,f_z,\eta)\frac{\partial T}{\partial x'}dx'+\sum_{\mu}\tilde{B}_{\mu }(x,f_y,f_z,\eta),
\end{equation}
where thermal conductivity $\kappa_x$ is given by 
\begin{equation}
\kappa_x(x,f_y,f_z,\eta) =\sum_{\mu_x>0, \mu_y, \mu_z}\kappa_{\mu x}\frac{e^{-\frac{1+i\Xi_{\mu}+if_y\Lambda_{\mu y}+if_z\Lambda_{\mu z}}{|\Lambda_{\mu x}|}x}}{(1+i\Xi_{\mu}+if_y\Lambda_{\mu y}+if_z\Lambda_{\mu z})|\Lambda_{\mu x}|}.
\end{equation}
$f_y$ and $f_z$ are the Fourier variables in the $y$ and $z$ directions, correspondingly, and $\tilde{B}_{\mu}(x,f_y,f_z,\eta)$ is the Fourier transform of $B_{\mu}(x,\rho,\zeta,\eta)$ with respect to $f_y$ and $f_z$. The exact expression of $\tilde{B}_{\mu}$ and a derivation of Eq.~(\ref{eq:heatflux_3D}) are given in Appendix \ref{app:3DDerivation}. In this case,  both temperature field and the inhomogeneous term have spatial and temporal dependence. Their dependence on the boundary conditions and heat source should be accounted for when extracting the intrinsic thermal conductivity from the observables such as total heat flux or an average temperature.



\section{Results}\label{sec:Results}

We now experimentally validate the generalized Fourier's law by comparing the predicted and measured surface thermal responses to incident heat fluxes in TDTR experiments. We consider a sample consisting of an aluminum film on a silicon substrate. Phonon dispersions for Al and Si and lifetimes for Si were obtained from first-principles using density functional theory (DFT).\cite{lindsay_abinitio_semiconductors_2013} We assumed a constant MFP for all modes in Al; the value $\Lambda_{Al} =$ 60 nm is chosen to yield a lattice thermal conductivity $\kappa \approx$ 123 Wm$^{-1}$K$^{-1}$ so that no size effects occur in the thin film. The justification of such an approach can be found in Ref.~\cite{Hua_2017}.

In a TDTR experiment, the in-plane directions are regarded as infinite, thus Fourier transforms in the $y$ and $z$ directions are justified. The cross-plane direction in the substrate layer is semi-infinite. Therefore, the cross-plane heat flux in the substrate is described by Eq.~(\ref{eq:heatflux_3D}) with $x \in [0, \infty)$. $P^{+}_{\mu}$ in Eq.~(\ref{eq:BoundaryConditions}) is determined by the interface conditions.\cite{Hua_2017}

In the diffusion regime, energy conservation, Fourier's law and the boundary conditions can fully describe a transport problem. In the nondiffusive regime, the replacement of Fourier's law is the generalized Fourier's law described in this work. If validated, this methodology allows the full prediction of the surface response for the wide variety of parameters employed in TDTR experiments, \emph{e.g.}, heating geometry, modulation frequency, and temperature.

\begin{figure*}[t!]
\centering
\includegraphics[scale = 0.45]{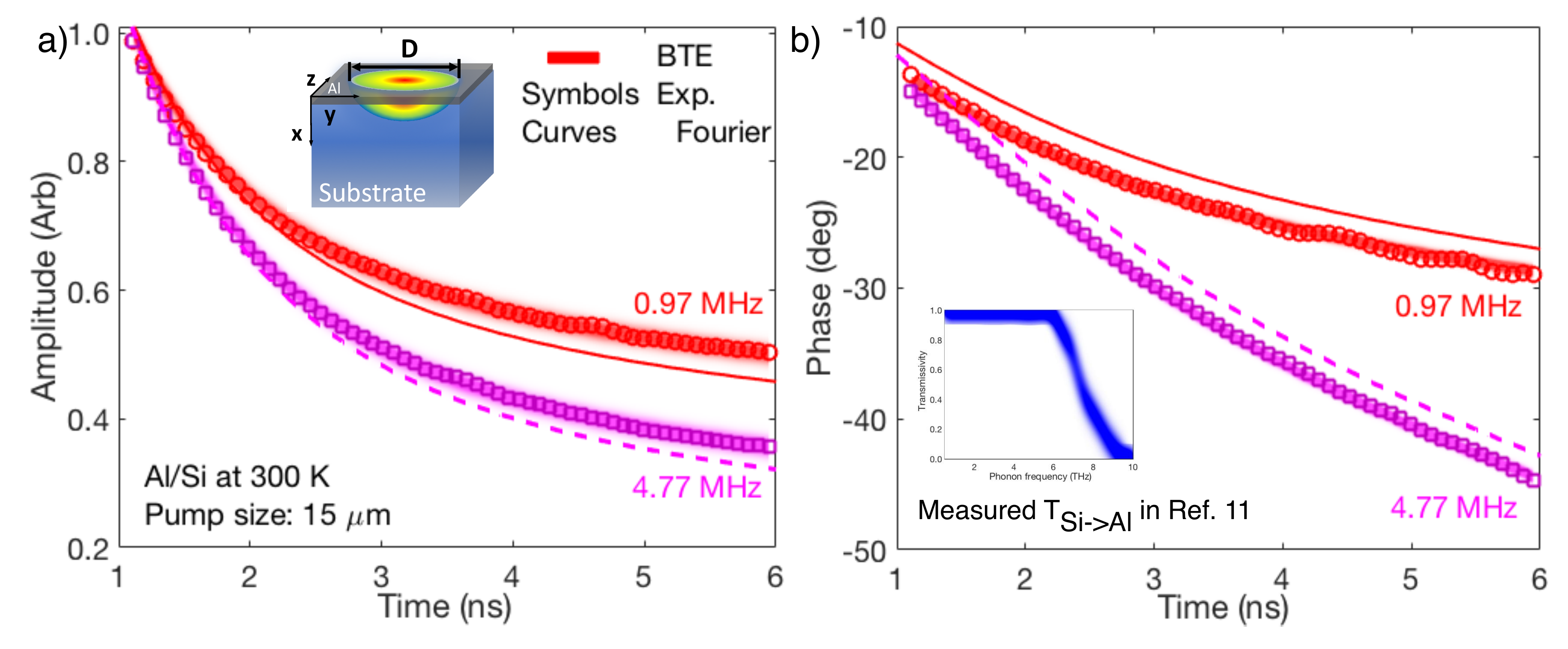}
\caption{ Experimental TDTR data (symbols) on the Al/Si sample with a pump beam size of 15 $\mu m$ at $T$ = 300 K for modulation frequencies of 0.97 MHz and 4.77 MHz along with the (a) amplitude and (b) phase compared with predictions from the generalized Fourier's law (shaded regions) and original Fourier's law (curves). The shaded regions around the PBE simulations correspond to the likelihood of the measured transmissivity possessing a particular value (darker area corresponds to more probable). Using the measured transmissivity profile from uniform heating,\cite{Hua_2017} the prediction from the generalized Fourier's law agrees well with the TDTR measurements, while the Fourier results overestimate the phase and underestimate the amplitude at later times. Inset in (a): schematic of the sample subject to a Gaussian beam heating. Inset in (b): the measured transmissivity of longitudinal phonons T$_{\text{Si} \rightarrow \text{Al}}(\omega)$ that is obtained under a uniform film heating in Ref. \cite{Hua_2017}.}
\label{Figure1}
\end{figure*}

To validate this methodology, we compare the calculated TDTR responses using the generalized Fourier's law with pump-size-dependent TDTR measurements on the same Al/Si sample measured in Ref. \cite{Hua_2017}, where the $1/e^2$ diameter $D$ of a Gaussian pump beam was varied between 5 to 60 $\mu$m at different temperatures. As the pump size decreases and becomes comparable to the thermal penetration depth along the cross-plane direction ($x$-axis of the schematic in Fig.~\ref{Figure1}(a)), in-plane thermal transport is no longer negligible and requires a three-dimensional transport description. $P^+_{\mu}(\xi_y,\xi_z,\eta)$ is determined from the interface condition, and for a given $\xi_y$, $\xi_z$, and $\eta$ it is determined by the spectral transmissivity of phonons as given in Ref. \cite{Hua_2017}. In the same work\cite{Hua_2017}, we have already used the PBE within the RTA to model the one-dimensional ($\xi_y = \xi_z = 0$) thermal transport in a TDTR experiment with a uniform film heating and developed a method to extract the spectral transmissivity profile of phonons from measurements. We also provided evidence that elastic transmission of phonons across an interface was the dominant energy transmission mechanism for materials with similar phonon frequencies. Therefore, the measured transmissivity profile at room temperature given in Ref.~\cite{Hua_2017} should be able to fully describe the interface conditions at other temperatures, and there are no adjustable parameters in the present simulations. 

\begin{figure*}[th!]
\centering
\includegraphics[scale = 0.45]{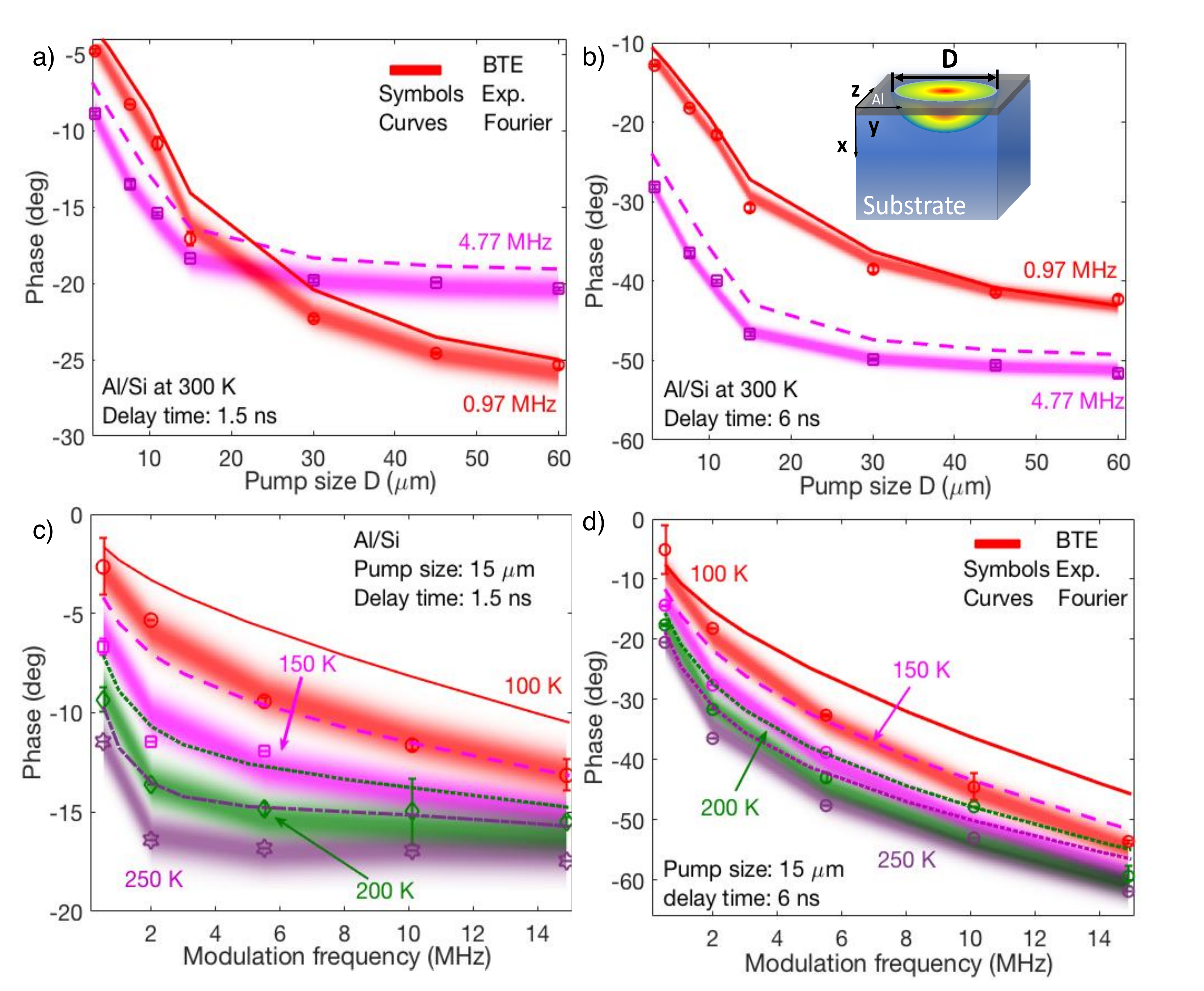}
\caption{Measured and predicted phases versus pump size at room temperature and a fixed delay time of (c) 1.5 ns and (d) 6 ns for modulation frequencies of 0.97 MHz and 4.77 MHz. At temperatures of 100, 150, 200, and 250 K, measured and predicted phases are plotted versus modulation frequency at a fixed delay time of (c) 1.5 ns and (d) 6 ns for a pump size of 15 $\mu$m. The symbols are TDTR data and the shaded regions are predictions. The curves show the prediction by Fourier's law with  temperature-dependent thermal conductivities from DFT and the interface conductance given by Eq.~(\ref{eq:InterfaceConductance}).}
\label{Figure2}
\end{figure*}

We compared the measured signals directly to predictions from the nonlocal transport governed by the generalized Fourier's law and the strictly diffusive transport governed by Fourier's law. To ensure a consistent comparison between the constitutive relations, the thermal conductivity of silicon is obtained using the same DFT calculations, and the interface conductance $G$ is given by\cite{zeng_nonequilibrium_2001}
\begin{equation}\label{eq:InterfaceConductance}
\frac{1}{G} = \frac{4}{\sum_{\mu} C^{Si}_{\mu } v^{Si}_{\mu } T_{\text{Si} \rightarrow \text{Al}}}-\frac{2}{\sum_{\mu} C^{Al}_{\mu } v^{Al}_{\mu }}-\frac{2}{\sum_{\mu} C^{Si}_{\mu } v^{Si}_{\mu }}
\end{equation}
where $T_{\text{Si} \rightarrow \text{Al}}$ is the phonon transmissivity from Si to Al. This expression was first derived by Chen and Zeng, which considers the non-equilibrium nature of phonon transport at the interface within the phonon transmissivity.\cite{zeng_nonequilibrium_2001} 

Figures~\ref{Figure1} (a) \& (b) show the total signal versus delay time with a pump size of 15 $\mu m$ at room temperature for the experiments and predictions from the generalized Fourier's law and original Fourier's law. As in Ref. \cite{Hua_2017}, the intensity of the shaded regions correspond to the likelihood of the measured transmissivity profile plotted in the inset of Fig.~\ref{Figure1}(b). A higher likelihood of a transmissivity profile is indicated by a higher intensity of the shaded area, and thus the PBE simulation using a transmissivity profile with higher likelihood better fits the experimentally measured TDTR signals. Excellent agreement between predictions from the generalized Fourier's law and experiments are observed. On the other hand, Fourier's law fails to accurately account for the experimental data, overestimating the phase and underestimating the amplitude after 2 ns in delay time. Note that the different transmissivity profiles in the inset of Fig.~\ref{Figure1}(b) give a value of interface conductance $G = 223 \pm 10$ Wm$^{-2}$K$^{-1}$ using Eq.~(\ref{eq:InterfaceConductance}), and this deviation in $G$ leads to uncertainties in the TDTR signals that are within the linewidth of the plotted curves. 

In Figs.~\ref{Figure2}(a) \& (b), comparisons of phases at different pump sizes between the generalized Fourier's law, original Fourier's law and experimental data are given at 300 K. In Figs.~\ref{Figure2}(c) \& (d), we compare the measured phase versus modulation frequency at a fixed pump size to predictions from the generalized Fourier's law and original Fourier's law at 100, 150, 200, 250 K. The data are given for two different delay times, 1.5 ns and 6 ns. The figure shows that predictions from the original Fourier's law do not reproduce the experimental results. The deviation of Fourier's law predictions from the experimental results becomes larger when the temperature decreases, modulation frequency increases, or pump size decreases, all indicating that non-diffusive effects increase with these changes. On the other hand, predictions from the generalized Fourier's law agree well with the experimental measurements for the various temperatures, modulation frequencies ,and pump sizes, indicating its validity to describe the nonlocal thermal transport in different regimes.

\section{Discussion}

\begin{figure*}[t!]
\centering
\includegraphics[scale = 0.5]{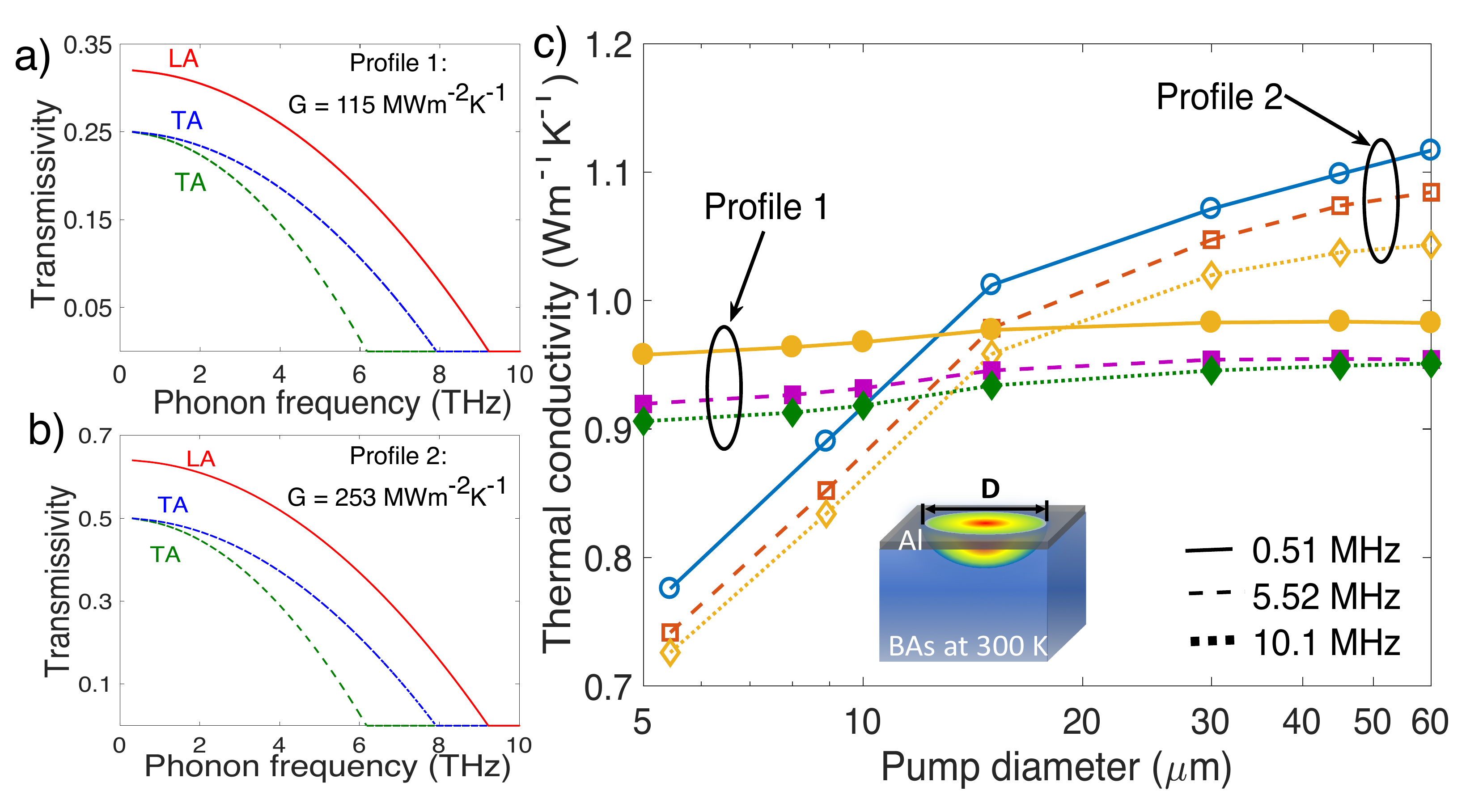}
\caption{Spectral transmissivity profiles from BAs to Al versus phonon frequency that give an interface conductance of (a) 115 MWm$^{-2}$K$^{-1}$ and (b) 253 MWm$^{-2}$K$^{-1}$ using Eq.~(\ref{eq:InterfaceConductance}). The profiles are used to generate the synthetic TDTR data. (c) Effective thermal conductivity versus pump size, obtained by fitting the synthetic TDTR data at the modulation frequencies of 0.51 (solid curves), 5.52 (dashed curves), and 10.1 (dotted curves) MHz, using the transmissivity profile (a) (solid symbols) and (b) (open symbols) to a diffusion model based on Fourier's law.  Both 3-phonon and 4-phonon scatterings are included in the DFT calculations of single crystalline BAs. The calculated bulk thermal conductivity of BAs is 1412 Wm$^{-1}$ K$^{-1}$.\cite{lindsay_BAs_DFT_2013,feng_four-phonon_2017} }
\label{fig:BAs_predictions}
\end{figure*}

All the above comparisons between the simulations and experiments with different heating geometries and at different temperatures provide evidence that the generalized Fourier's law is an appropriate replacement of Fourier's law in the nondiffusive regime. We now use this formalism to examine TDTR measurements on boron arsenide (BAs).

Boron arsenide has recently attracted attention because of its ultra-high thermal conductivity determined from measurements based on the TDTR technique and reported by several groups.\cite{Lieaat8982,Kangeaat5522,Tianeaat7932} Moreover, pump-size-dependent measurements have also been conducted in an attempt to access information of the phonon MFPs in BAs.\cite{Kangeaat5522} The thermal conductivity measurements are based on interpreting the raw TDTR data as fit to a diffusion model based on Fourier's law with thermal conductivity of BAs and interface conductance between Al and BAs as two fitting parameters. 

However, due to the presence of phonons with long MFPs compared to the TDTR thermal length scale, Fourier's law is no longer valid at the length scales probed by TDTR. As predicted by DFT-based PBE calculations\cite{lindsay_BAs_DFT_2013,feng_four-phonon_2017}, more than 70\% of phonons in single crystalline BAs have mean free paths longer than 1 $\mu m$. Therefore, properly interpreting the data requires using the generalized Fourier's law. 

Equation~(\ref{eq:HeatFlux_v2}) needs close examination to understand the microscopic information contained in the surface temperature responses measured in experiments. In a two-layer structure like the one used in TDTR, the second term in Eq.~(\ref{eq:HeatFlux_v2}) does not vanish and has a non-local nature through the source term. This nonlocality implies that even though only the transient temperature at the metal surface is observed, the measurement contains information from the entire domain. We have demonstrated in Ref. \cite{Hua_2018} that the spectral distribution of the source term alters the surface temperature response. In other words, even though the first term in Eq.~(\ref{eq:HeatFlux_v2}) remains the same, the observable at the surface is altered by the inhomogeneous source term originating from the interface. 

To illustrate this point, we choose two transmissivity profiles T$_{\text{BAs}\rightarrow \text{Al}}$ as shown in Figs.~\ref{fig:BAs_predictions} (a) \& (b). The profiles share a similar dependence on phonon frequency but with a different magnitude. The nominal interface conductance is calculated to be $115$ and $253$ MWm$^{-2}$K$^{-1}$, respectively, using Eq.~(\ref{eq:InterfaceConductance}). Along with the \emph{ab initio} properties of BAs at room temperature, we calculate the TDTR responses at the Al surface with different pump sizes at three modulation frequencies. The calculated TDTR responses are then fit to the typical diffusion model based on Fourier's law to extract the effective thermal conductivity, as was performed in Refs \cite{Lieaat8982,Kangeaat5522,Tianeaat7932}. 

The results are shown in Fig.~\ref{fig:BAs_predictions}(c). The key observation from Fig.~\ref{fig:BAs_predictions}(c) is that the effects of modulation frequency and pump size on the effective thermal conductivity depend on the transmissivity profiles. A decrease in the effective thermal conductivity is observed in both profiles as the pump size decreases or the modulation frequency increases. However, the magnitude of the reduction and the absolute value compared to the bulk value depend on the transmissivity. While the effective thermal conductivity seems to be approaching the bulk value at a large pump size and low modulation frequency using profile 1, the effective thermal conductivity using profile 2 exceeds the bulk value under the same conditions. The reduction of the effective thermal conductivity using profile 1 as pump size decreases is less than 5\% at a given modulation frequency, while the reduction using profile 2 can be as much as 40\%. 

Our calculations therefore indicate that in the nondiffusive regime, simply interpreting measurements from a method such as TDTR using Fourier's law with a modified thermal conductivity may yield incorrect measurements. Not only is Fourier's law unable to describe the nonlocal nature of thermal conductivity, but it also does not include the effects of inhomogeneous terms. Therefore, when interpreting TDTR measurements of high thermal conductivity materials, Fourier's law is not the appropriate constitutive relation. In contrast, we have provided experimental evidence that the generalized Fourier's law, Eq.~(\ref{eq:HeatFlux_v2}), accurately describes thermal transport in the non-diffusive regime. 

\section{Conclusions}

In summary, we derived a generalized Fourier's law using the Peierls-Boltzmann equation under the relaxation time approximation. The new constitutive relation consists of two parts, a convolution between the temperature gradient and a space- and time-dependent thermal conductivity, and an inhomogeneous term determined from boundary conditions and heat sources. By comparing predictions from this new constitution law to a series of time-domain thermorflectance measurements in the nondiffusive regime, we provide experimental evidence that the generalized Fourier's law more accurately describes thermal transport in a range of transport regimes. We also show that interpreting nonlocal thermal transport using Fourier's law can lead to erroneous interpretation of measured observables. To correctly extract microscopic phonon information from the observation of nonlocal thermal transport, it is necessary to separate the inhomogeneous contribution from the nonlocal thermal conductivity based on the generalized Fourier's law developed here. 

\section{Acknowledgements}

C. H. and L. L. acknowledge support from the Laboratory Directed Research and Development Program of Oak Ridge National Laboratory, managed by UT-Battelle, LLC, for the U.S. Department of Energy. A. J. M. acknowledges support from the National Science Foundation under Grant No. CBET CAREER 1254213. This research used resources of the National Energy Research Scientific Computing Center (NERSC), a U.S. Department of Energy Office of Science User Facility operated under Contract No. DE-AC02-05CH11231.

\appendix

\section{Derivation of Eq.~(\ref{eq:TGHeatFlux})}\label{app:TGDerivation}

In a one-dimensional (1D) problem, Eq.~(\ref{eq:HeatFlux_v2}) becomes
\begin{equation}\label{eq:ModeSpecificHeatFluxTG_app}
\tilde{q}_{\mu x} = - \int_{\Gamma}C_{\mu}v_{\mu x}\Lambda_{\mu x}\frac{e^{-\alpha_{\mu}\left|\frac{x-x'}{\Lambda_{\mu x}}\right|}}{\alpha_{\mu}|\Lambda_{\mu x}|}\frac{\partial \tilde{T}}{\partial x'}dx '+\int_{\Gamma}\tilde{Q}_{\mu}(x')e^{-\alpha_{\mu}\left|\frac{x-x'}{\Lambda_{\mu x}}\right|}dx' ,
\end{equation}
where 
$$
\Gamma \in \left\{ \begin{array}{rl}
 [-\infty, \xi)&\mbox{ if $v_{\mu\xi}>0$} \\
 \left(\xi , \infty \right] &\mbox{ if $v_{\mu\xi}<0$}
\end{array} \right. .
$$

In a 1D transient grating experiment, both temperature profile and mode-specific heat input rate have a spatial dependence of $e^{i\beta x}$, i.e., $\tilde{T}(x,\eta) = e^{i\beta x}\tilde{T}(\eta)$ and $\tilde{Q}_{\mu}(x) = Q_{\mu}\delta^{-1} e^{i\beta x}$. Summing Eq.~(\ref{eq:ModeSpecificHeatFluxTG_app}) over all the phonon modes and using the symmetry of $v_{\mu x}$ about the center of the Brillouin zone, i.e., $v_{\mu x} = -v_{-\mu x}$, the total heat flux is expressed as
\begin{equation}
\tilde{q}_{ x} = -i\beta\Delta \tilde{T}(\eta)\sum_{\mu_x >0} \int^{\infty}_{-\infty}C_{\mu}v_{\mu x}\Lambda_{\mu x}\frac{e^{i\beta x'}e^{-\alpha_{\mu}\left|\frac{x-x'}{\Lambda_{\mu x}}\right|}}{\alpha_{\mu}|\Lambda_{\mu x}|}dx '+\sum_{\mu_x >0}\int^{\infty}_{-\infty}\frac{Q_{\mu}}{\delta}e^{i\beta x'}e^{-\alpha_{\mu}\left|\frac{x-x'}{\Lambda_{\mu x}}\right|}dx'.
\end{equation}
Now define $y = x'-x$. Then the above equation becomes:
\begin{eqnarray}\label{eq:TGHeatFlux_app}
\tilde{q}_{x} &=& -i\beta\Delta \tilde{T}(\eta) e^{i\beta x}\sum_{\mu_x >0}\int^{\infty}_{-\infty}\sum_{\mu x >0}C_{\mu}v_{\mu x}\Lambda_{\mu x}\frac{e^{i\beta y}e^{-\alpha_{\mu}\left|\frac{y}{\Lambda_{\mu x}}\right|}}{\alpha_{\mu}|\Lambda_{\mu x}|}dy+e^{i\beta x}\sum_{\mu_x >0}\int^{\infty}_{-\infty}\frac{Q_{\mu}}{\delta}e^{i\beta y}e^{-\alpha_{\mu}\left|\frac{y}{\Lambda_{\mu x}}\right|}dy \nonumber \\
&=&  - i\beta\tilde{T}(\eta)e^{i\beta x}\sum_{\mu_x>0}\frac{\kappa_{\mu x}}{\alpha_{\mu}^2+\Lambda_{\mu x}^2\beta^2}+e^{i\beta x}\sum_{\mu_x >0}\frac{Q_{\mu}}{\delta}\frac{\alpha_{\mu}\Lambda_{\mu x}}{\alpha_{\mu}^2+\Lambda_{\mu x}^2\beta^2}.
\end{eqnarray}

\section{Derivation of Eq.~(\ref{eq:heatflux_3D})}\label{app:3DDerivation}

When the $y$ and $z$ directions can be regarded as infinite, the analytical mapping to the Cartesian coordinates can be completed via Fourier transform in $y$ and $z$.  To show it , we first define $g(x,y,z) = f(\xi ,\rho ,\zeta )$ with the coordinate transform given by Eq.~(\ref{eq:newvariables}). $G$ and $F$ are the functions after Fourier transform in $y$ and $z$. Using the affine theorem of two-dimensional Fourier transform, we obtain
\begin{equation}\label{eq:AffineTheorem}
G(x,f_y,f_z) = e^{-i\left(f_y\frac{\Lambda_{\mu y}}{\Lambda_{\mu x}}+f_z\frac{\Lambda_{\mu y}}{\Lambda_{\mu x}}\right)x}F(x,-f_y\frac{\Lambda_{\mu}}{\Lambda_{\mu x}},-f_z\frac{\Lambda_{\mu}}{\Lambda_{\mu x}})\frac{\Lambda^2_{\mu}}{\Lambda^2_{\mu x}},
\end{equation}
where $f_y$ and $f_z$ are the Fourier variables in the $y$ and $z$ directions, respectively. 

Applying Eq.~(\ref{eq:AffineTheorem}) to Eq.~(\ref{eq:HeatFlux}) gives
\begin{eqnarray}\label{eq:ModeHF_app}
\tilde{q}_{\mu\xi}= P^{\ast}_{\mu}(f_y,f_z,\eta)v_{\mu x}e^{-\frac{1+i\gamma_{\mu}}{\Lambda_{\mu x}}x} &+& sgn(v_{\mu x})\int_{\Gamma}\tilde{Q}_{\mu}(x,f_y,f_z,\eta)e^{-\frac{1+i\gamma_{\mu}}{|\Lambda_{\mu x}| }|x-x '|}dx ' \nonumber \\
&+& \int_{\Gamma}\frac{C_{\mu}v_{\mu x}}{|\Lambda_{\mu x}|} \Delta \tilde{T}(x ',f_y,f_z,\eta)e^{-\frac{1+i\gamma_{\mu}}{|\Lambda_{\mu x}| }|x-x'|}dx ',
\end{eqnarray}
where $\gamma_{\mu} =\eta\tau_{\mu}+f_y\Lambda_{\mu y}+f_z\Lambda_{z}$. Note that 
\begin{eqnarray}
\int\int P(\rho,\zeta,\eta)e^{if_yy+if_zz}dydz &=& P_{\mu}(-f_y\frac{\Lambda_{\mu}}{\Lambda_{\mu x}},-f_z\frac{\Lambda_{\mu}}{\Lambda_{\mu x}},\eta)\frac{\Lambda^2_{\mu}}{\Lambda^2_{\mu x}}e^{-i\left(f_y\frac{\Lambda_{\mu y}}{\Lambda_{\mu x}}+f_z\frac{\Lambda_{\mu z}}{\Lambda_{\mu x}}\right)x} \nonumber \\
&=& P^{\ast}_{\mu}(f_y,f_z,\eta)e^{-i\left(f_y\frac{\Lambda_{\mu y}}{\Lambda_{\mu x}}+f_z\frac{\Lambda_{\mu z}}{\Lambda_{\mu x}}\right)x}.
\end{eqnarray}

Applying integration by parts to the third term in Eq.~(\ref{eq:ModeHF_app}) and summing up all the phonon modes gives
\begin{equation}\label{eq:heatflux_3D_app}
\tilde{q}_{x}(x,f_y,f_z,\eta) = -\int^{L_2}_{L_1}\kappa_x(x-x',f_y,f_z,\eta)\frac{\partial T}{\partial x'}dx'+\tilde{B}(x,f_y,f_z,\eta),
\end{equation}
where thermal conductivity $\kappa_x$ is given by 
\begin{equation}
\kappa_x(x,f_y,f_z,\eta) =\sum_{\mu_x>0, \mu_y, \mu_z}\kappa_{\mu x}\frac{e^{-\frac{1+i\gamma_{\mu}}{|\Lambda_{\mu x}|}x}}{(1+i\gamma_{\mu})|\Lambda_{\mu x}|},
\end{equation}
and
\begin{eqnarray}
\tilde{B}(x,f_y,f_z,\eta) &=& \sum_{\mu}P^{\ast}_{\mu}(f_y,f_z,\eta)v_{\mu x}e^{-\frac{1+i\gamma_{\mu}}{\Lambda_{\mu x}}x} \nonumber \\
&+& \sum_{\mu_x >0,\mu_y,\mu_z}\frac{C_{\mu}|v_{\mu x}|}{1+i\gamma_{\mu}}\left[\Delta T(L_2) e^{-\frac{1+i\gamma_{\mu}}{\Lambda_{\mu x}}(L_2-x)}-\Delta T(L_1) e^{-\frac{1+i\gamma_{\mu}}{\Lambda_{\mu x}}(x-L_1)}\right] \nonumber \\
&+& \sum_{\mu_x >0,\mu_y,\mu_z}\frac{|\Lambda_{\mu x}|}{1+i\gamma_{\mu}}\left[Q_{\mu}(L_2) e^{-\frac{1+i\gamma_{\mu}}{\Lambda_{\mu x}}(L_2-x)}-Q_{\mu}(L_1) e^{-\frac{1+i\gamma_{\mu}}{\Lambda_{\mu x}}(x-L_1)}\right] \nonumber \\
&-& \sum_{\mu_x >0,\mu_y,\mu_z}\frac{|\Lambda_{\mu x}|}{1+i\gamma_{\mu}}\int^{L_2}_{L_1}\frac{\partial Q_{\mu}}{\partial x'}e^{-\frac{1+i\gamma_{\mu}}{|\Lambda_{\mu x}|}|x'-x|}dx'.
\end{eqnarray}

\clearpage


\end{document}